\documentclass[aps,prl,twocolumn,superscriptaddress]{revtex4-1}   
\usepackage{graphicx,amssymb}
\usepackage{color,ulem}
\usepackage{bm}   

\newcommand{\eqref}[1]{(\ref{#1})}

\begin{document}

\title{Quantum trajectory analysis of single microwave photon detection by nanocalorimetry}
\author{Bayan Karimi}
\affiliation{QTF Centre of Excellence, Department of Applied Physics, Aalto University School of Science, P.O. Box 13500, 00076 Aalto, Finland}
\author{Jukka P. Pekola}
\affiliation{QTF Centre of Excellence, Department of Applied Physics, Aalto University School of Science, P.O. Box 13500, 00076 Aalto, Finland}

\date{\today}

\begin{abstract}
We apply quantum trajectory techniques to analyze a realistic set-up of a superconducting qubit coupled to a heat bath formed by a resistor, a system that yields explicit expressions of the relevant transition rates to be used in the analysis. We discuss the main characteristics of the jump trajectories and relate them to the expected outcomes ("clicks") of a fluorescence measurement using the resistor as a nanocalorimeter. As the main practical outcome we present a model that predicts the time-domain response of a realistic calorimeter subject to single microwave photons, incorporating the intrinsic noise due to the fundamental thermal fluctuations of the absorber and finite bandwidth of a thermometer.
\end{abstract}
\maketitle
Quantum trajectories provide a way to predict the stochastic behaviour of an open quantum system experiencing the subtle influence of the environment via a non-Hermitian Hamiltonian, and jumps between eigenstates. Initially developed about 30 years ago as a computational aid~\cite{KlausM, KlausM2, Zoller, Carmichael}, the trajectories are nowadays routinely used for interpretation of experiments even in modern macroscopic quantum systems~\cite{Wineland,Vijay, Murch, Rossi, Brun,molmer2, Minev}. For instance, in the currently active field of quantum thermodynamics, quantum trajectories provide an invaluable tool to describe the stochastic thermodynamics properties of open quantum systems~\cite{Jukka,Leggio,cyril,Lutz,utsumi,donvil}. In this paper we present an analysis of an archetypical basic set-up: a two-level system (qubit) coupled to a heat bath. In particular, we take a concrete system of a solid-state superconducting qubit~\cite{oliver} and resistive environment forming an equilibrium heat bath, which is readily realizable experimentally~\cite{valve, Mikko}. We focus here on the expected outcomes of a fluorescence measurement based on observing emitted and absorbed microwave photons by a nanocalorimeter that presents a circuit realization of a photoreceiver discussed in general terms, e.g. in~\cite{hw}. We demonstrate that the common interpretation of the outcome of a projective measurement ("collapse") is consistent with the quantum jump trajectories. We present a stochastic simulation of the output of this detector in the presence of qubit-calorimeter interaction and coupling of the calorimeter to the phonon heat bath including thermal noise on the detector. This analysis illustrates the feasibility of such an experiment under realistic conditions, and its potential to detect not only the arrival times but also the energies of the quanta in a continuous measurement in the challenging regime of microwave photons. 
\begin{figure}[h!!!]
	\includegraphics[width=7.8 cm]{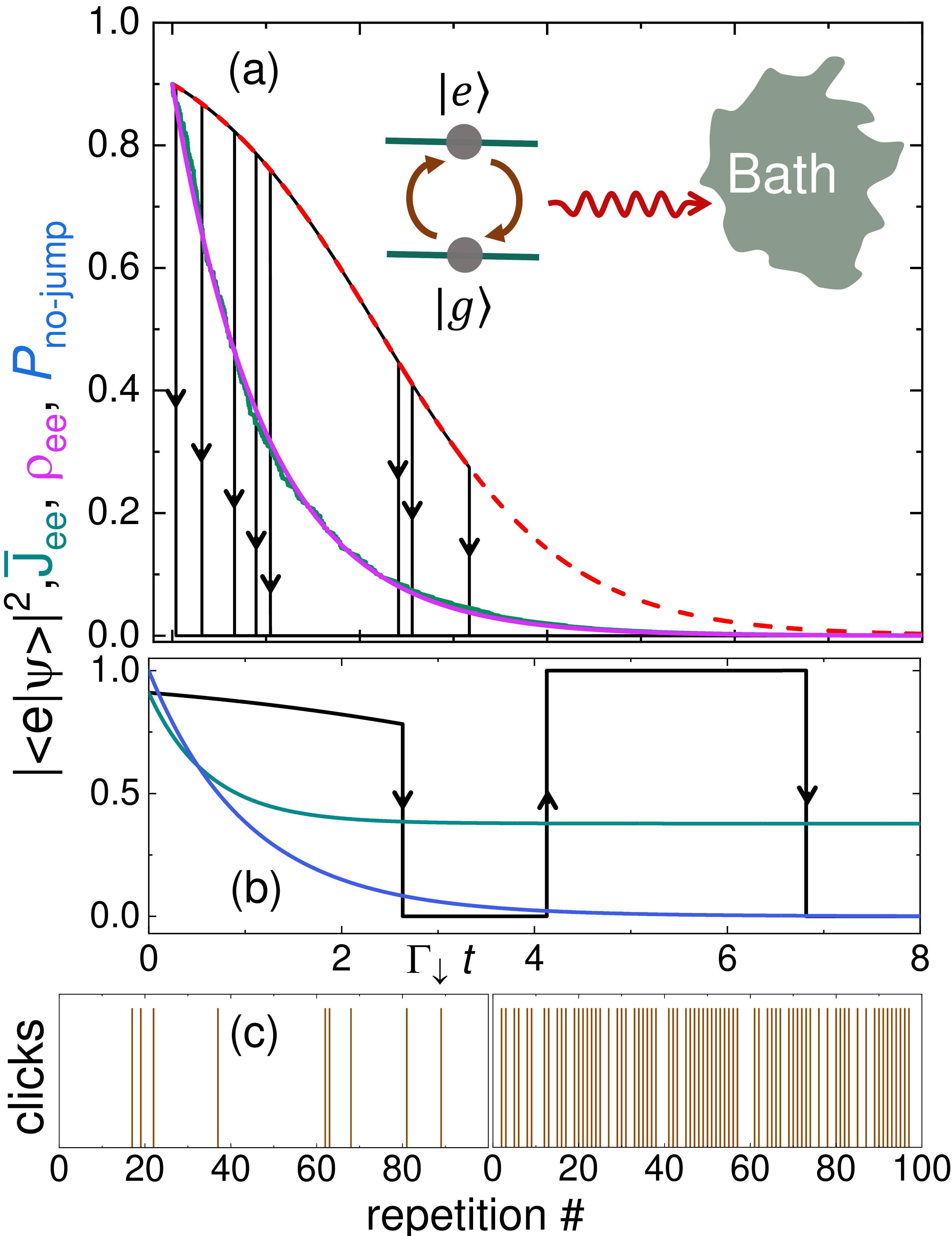}
	\caption{Two level system (qubit) coupled to a heat bath, shown in the inset. (a) Time evolution of the qubit initially prepared in the state $|\psi(0)\rangle$ when coupled to zero temperature bath. We assume $|\langle e|\psi(0)\rangle|^2=0.9$. The dashed line indicates the time dependence of $|\langle e|\psi(t)\rangle|^2$ for no-jump trajectories. In general $|\langle e|\psi(t)\rangle|^2$ follows the dashed line until the stochastic jump occurs. We also present overlapping $\rho_{ee}(t)$ and $\bar{J}_{ee}(t)$, an estimate of $J_{ee}(t)= \overline{|\langle e|\psi(t)\rangle|^2}$, by averaging 100 trajectories. (b) Same as (a) at finite temperature $\beta\hbar\omega_{\rm Q}=0.5$. $\bar{J}_{ee}(t)$ is averaged in this case over $10^5$ realizations. $P_{\rm no-jump}(t)$ (Eq. \eqref{pnojumpT}) shows double-exponential decay. (c) Jumps into the ground state in a sequence of 100 repetitions preparing the qubit initially in the state $|\psi(0)\rangle$ where in left (right) panel $|\langle g|\psi(0)\rangle|^2=0.9~(0.3)$ and $|\langle e|\psi(0)\rangle|^2=0.1~(0.7)$. In both panels temperature is zero.}
	\label{qubitbath}
\end{figure}

We consider a qubit coupled to a heat bath 
as schematically shown in the inset of Fig.~\ref{qubitbath}a. The stochastic wave function of this system,
\begin{equation}\label{wave} |\psi(t)\rangle=a(t)|g\rangle+b(t)|e\rangle,
\end{equation}
is written in the basis of the ground $|g\rangle$ and excited $|e\rangle$ states. The non-Hermitian Hamiltonian of the system is

\begin{equation}\label{totalHamiltonian2}
H=H_{\rm S}-\frac{i\hbar}{2}\Gamma_\downarrow|e\rangle \langle e|-\frac{i\hbar}{2}\Gamma_\uparrow|g\rangle \langle g|.
\end{equation}
Here $H_{\rm S}=-\frac{1}{2}\hbar\omega_{\rm Q}\sigma_z$ in the $\{ |g\rangle, |e\rangle\}$ basis is the bare Hamiltonian of the qubit with $\hbar\omega_{\rm Q}$ the energy level spacing of it and $\sigma_z$ the z-component of the Pauli matrix. $\Gamma_{\uparrow,\downarrow}$ are the excitation and relaxation rates of the qubit, whose precise forms will be obtained later via the master equation (ME). When no jump occurs the wave function evolves as 
$|\psi^{(0)}(t+dt)\rangle=\frac{1}{\sqrt{1-dp}}(1-\frac{i dt}{\hbar} H)|\psi(t)\rangle$, with
$dp=dp_\uparrow+dp_\downarrow$ where $dp_\downarrow=\Gamma_\downarrow |b(t)|^2dt$ is the probability that the jump occurs down to $|g\rangle$ in a short time interval $dt$, and corespondingly $dp_\uparrow=\Gamma_\uparrow |a(t)|^2dt$ is the probability to jump up to $|e\rangle$~\cite{KlausM}. For no-jump trajectories we thus have $\dot{a}(t)=\frac{1}{2}\Delta\Gamma a(t)|b(t)|^2$ and $\dot{b}(t)=-\frac{1}{2}\Delta\Gamma b(t)|a(t)|^2$ 
where $\Delta\Gamma\equiv \Gamma_\downarrow-\Gamma_\uparrow$. 

Our exemplary protocol drives the two-level system initially into the superposition $|\psi(0)\rangle=a(0)|g\rangle +b(0)|e\rangle$ whereafter it is let to evolve freely though coupled to the bath. Based on the equations above we have
\begin{eqnarray}\label{prob2}
&&|a(t)|^2={|a(0)|^2e^{-\Gamma_\uparrow t}}/P_{\rm no-jump}(t)\nonumber\\&&|b(t)|^2={|b(0)|^2e^{-\Gamma_\downarrow t}}/P_{\rm no-jump}(t).
\end{eqnarray}
Here $P_{\rm no-jump}(t)=e^{-\int_0^t(\Gamma_\downarrow |b(t')|^2 +\Gamma_\uparrow |a(t')|^2) dt'}$ is the probability that no jump occurs until time $t$,
\begin{equation}\label{pnojumpT}
P_{\rm no-jump}(t)=|a(0)|^2e^{-\Gamma_\uparrow t}+|b(0)|^2e^{-\Gamma_\downarrow t}.
\end{equation}
It satisfies the two conditions $P_{\rm no-jump}(0)=1$ and $P_{\rm no-jump}(\infty)=0$, the latter meaning that jump takes place eventually as shown in Fig.~\ref{qubitbath}b. 

In literature on quantum optics, the jumps are typically given by dissipators related to Lindblad type master equations~\cite{KlausM2,BP}, without explicit relation to the concrete bath. On the contrary, here in our system, we can make reference to the actual set-up, and obtain the relevant rates $\Gamma_{\uparrow,\downarrow}$ given by the circuit and the well-defined bath that it is coupled to. In order to find the expression for these transition rates and the population of the eigenstates in time, we derive the ME for this system. In the standard weak-coupling theory, the total Hamiltonian can be written as
\begin{equation} \label{h1}
H_{\rm tot} =H_{\rm S}+ V(t)+H_{\rm B},
\end{equation}
where $H_{\rm B}$ is the Hamiltonian of the bath and $V(t)$ is the coupling energy between the system and the bath. 

\begin{figure}
	\includegraphics[width=7cm,height=3cm,keepaspectratio]{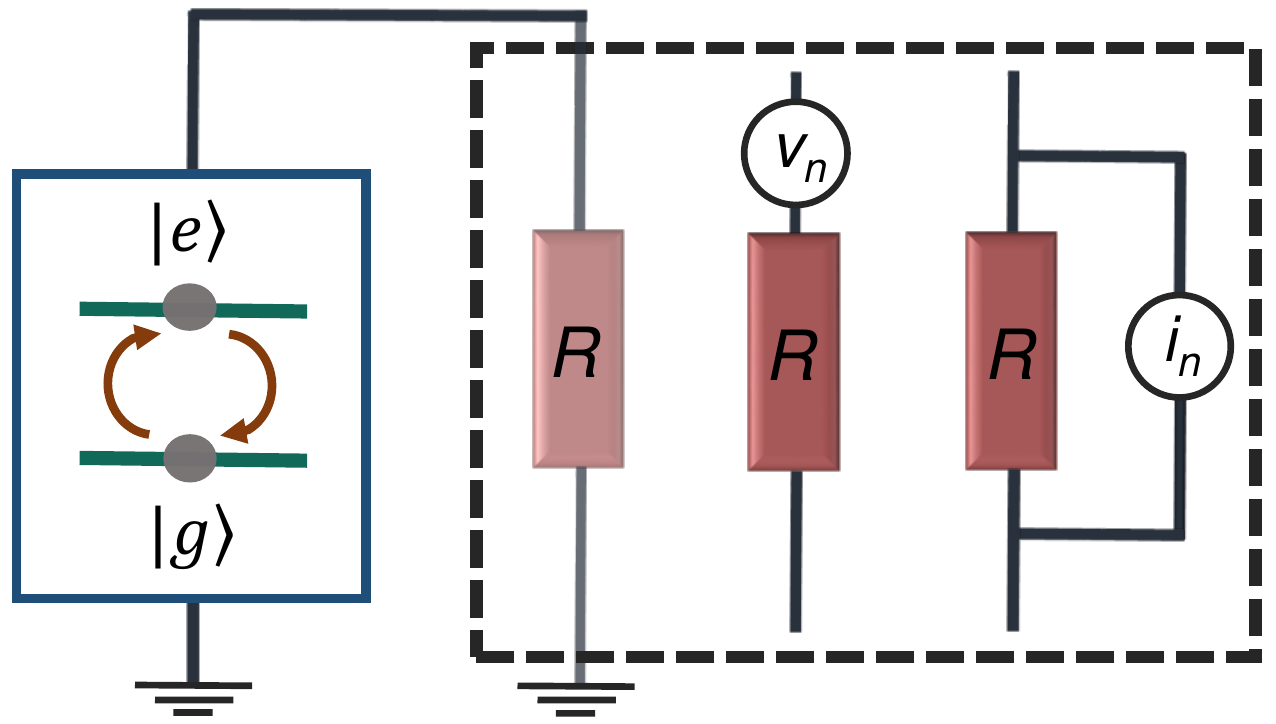}
	\caption{A qubit coupled to a heat bath shown as a resistor $R$. Depending on the design of the circuit and operating regime, we consider it as a voltage or current source of thermal noise.}
	\label{circuit}
\end{figure}

For the perturbation, we assume that it is produced by a resistor $R$ (Fig.~\ref{circuit}) forming the immediate bath of the qubit, and we take linear coupling as $V(t)=AX_n(t)$, where $A$ is an operator of the system and $X_n(t)$ is the noise of the resistor. As in this figure depending on the configuration (current or voltage biasing) one can choose either $V(t)=\Phi i_n(t)$ or $V(t)=qv_n(t)$, where $\Phi~(q)$ is the phase (charge) operator, and $i_n(t)~(v_n(t))$ denotes the current (voltage) noise. Without loss of generality we take the first option: the final results will be identical for the two possible choices with the proper definition of the quality factor of the system. We have for the system density matrix $\rho(t)$ in the interaction picture, 
\begin{equation} \label{h9}
\dot \rho(t)= -\frac{1}{\hbar^2} \int_{-\infty}^t {\rm Tr_{\rm B}} \Big\{ \big [[\rho(t)\otimes\rho_{\rm B}, V_{\rm I}(t')],V_{\rm I}(t)\big ]\Big\}dt', 
\end{equation}
where $V_{\rm I}(t)=e^{iH_{\rm S}t/\hbar}V(t)e^{-iH_{\rm S}t/\hbar}$, $\rho_{\rm B}$ is the density matrix of the bath and ${\rm Tr_{\rm B}}$ denotes trace over it. The diagonal and off-diagonal elements of the master equation, $\rho_{gg}$ and $\rho_{ge}$, respectively, are then given by 
\begin{equation}\label{rhoggrhoge}
\dot \rho_{gg}(t)=-\Gamma_\Sigma \rho_{gg}(t)+\Gamma_\downarrow,~\dot \rho_{ge}(t)=-\frac{1}{2}\Gamma_\Sigma \rho_{ge}(t),
\end{equation}
with $\Gamma_\Sigma=\Gamma_\downarrow+\Gamma_\uparrow$. In accordance with this analysis the rates obey the Fermi golden rule expressions 
\begin{equation}\label{rates}
\Gamma_{\downarrow,\uparrow}=\frac{1}{\hbar^2}|\langle g|\Phi|e\rangle|^2 S_i(\pm \omega_{\rm Q}).
\end{equation}
Here the noise spectral density of current is given by $S_i(\pm \omega)={2R^{-1}\hbar\omega}/{(1-e^{-\beta\hbar\omega})}$ at angular frequency $\omega$. Note that this noise that governs the transition rates is determined by the temperature $T=k_{\rm B}/\beta$ of the absorber, which may vary in time since we assume that this absorber is a mesoscopic bath coupled to the real "superbath"~\cite{Campisi} at a constant temperature $T_0$. For a qubit that can be approximated by an $LC$ resonator, we can express the phase operator as $\Phi=\sqrt{\hbar Z_0/2}(\hat{a}+\hat{a}^\dagger)$ where $Z_0=\sqrt{L_{\rm J}/C_{\rm J}}$ with $L_{\rm J}$ and $C_{\rm J}$ the (Josephson) inductance and capacitance of the qubit, and $\hat{a}=|g\rangle \langle e|$. We obtain for the transition rates 
\begin{eqnarray}\label{transitions}
\Gamma_\downarrow=\frac{1}{Q}\frac{\omega_{\rm Q}}{1-e^{-\beta \hbar \omega_{\rm Q}}},~~\Gamma_\uparrow=\frac{1}{Q}\frac{\omega_{\rm Q}}{e^{\beta \hbar \omega_{\rm Q}-1}},
\end{eqnarray}
where the dependence on the specific set-up comes only via the quality factor $Q=Z_0/R$. The transition rates obey the detailed balance condition, $\Gamma_\uparrow = e^{-\beta \hbar\omega_{\rm Q}} \Gamma_\downarrow$. To further connect the results with a concrete circuit, we note that the quality factor relates to the standard $T_1$ relaxation time of the qubit by $T_1=Q/\omega_{\rm Q}$ at low temperature~\cite{oliver}.
As a sanity check, we return to the stochastic wavefunction and calculate the quantity $J(t)\equiv \overline{|\psi(t+dt)\rangle \langle\psi(t+dt)|}$, the average over many trajectories, which is expected to mimic the density matrix. For this system
\begin{eqnarray}\label{Jbar2}
J(t)=&&(1-dp)|\psi^{(0)}(t+dt)\rangle \langle \psi^{(0)}(t+dt)|\nonumber\\&&+dp_\downarrow |g\rangle \langle g|+dp_\uparrow |e\rangle \langle e|\nonumber.
\end{eqnarray}
We then have by a straightforward calculation
\begin{equation}\label{diag}
\dot J_{gg}(t)=-\Gamma_\Sigma J_{gg}(t)+\Gamma_\downarrow,~\dot J_{ge}(t)=-\frac{1}{2}\Gamma_\Sigma J_{ge}(t).
\end{equation}  
As expected, Eqs.~\eqref{rhoggrhoge} and \eqref{diag} are identical by interchanging $J$ for $\rho$. 

The main panels of Fig.~\ref{qubitbath} summarize the results presented up to now with given parameters. In Fig.~\ref{qubitbath}a, with $T=0$, the dashed red line presents the no-jump evolution of $|b(t)|^2$ and the abrupt transitions down indicate the stochastic quantum jumps to the ground state according to Monte-Carlo simulations. In these simulations the jump probabilities are determined by $dp_\downarrow$ as described above. The two other overlapping lines are from averaging $|\langle e|\psi(t)\rangle|^2$ over 100 trajectories, yielding an estimate of $J_{ee}(t)$, and $\rho_{ee}(t)$ from the presented ME. In the b panel similar quantities (same colours) are shown at a finite bath temperature, demonstrating jumps also to the excited state, yielding a non-vanishing $\rho_{ee}(t)$ when $t\rightarrow \infty$. Let us next apply the obtained concrete framework to the actual calorimetric measurement.

{\it Bath (absorber) as the measuring device (detector):}
The temperature of the absorber $T$ is the quantity that we monitor (fluorescence measurement). We assume that the detector is able to tell whether a photon is absorbed or emitted based on the temperature change due to such an event. This is possible if the resistor is a finite-size absorber that is only weakly coupled to an infinite bath. From the practical measurement point of view we adopt the philosophy that following the "state" of the many-body detector, the resistor, in our case by measuring its temperature by a local non-invasive thermometer~\cite{ZBA}, we do not influence the stochastic trajectories of the system, as argued in~\cite{hw}. 
Also, we ignore the fact that the transition rates are influenced by the variations of the instantaneous $T$ of the absorber. We focus now on the measurement of the first, i.e., the "guardian" photon after the two-level system is prepared in the general superposition of Eq. \eqref{wave} at $t=0$. The probability that this photon is absorbed by the detector ("click-up"), corresponding to the transition $\downarrow$ of the qubit, is given by
\begin{equation}\label{measu.}
P_{\rm click-up}=\int_{0}^{\infty} P_{\rm no-jump}(t')\frac{dp_\downarrow(t')}{dt'}dt'=|\langle e|\psi(0)\rangle|^2.
\end{equation}
By the same argument we would then obtain that the first photon is emitted by the detector with the probability $P_{\rm click-down}=|\langle g|\psi(0)\rangle|^2$. We note the following: (i) These results hold for any temperature of the absorber. (ii) The arrival time of the guardian photon is stochastic. (iii) It is natural that only the first photon plays a role here, since the next one probes the state of the system after the previous jump and so on.
Figure~\ref{qubitbath}c illustrates how the above principle is realized in a numerical experiment. We show results of repeated protocols with two different initial states (100 realizations each), where $|\langle g|\psi(0)\rangle|^2=0.9$ and $0.3$ on the left and right, respectively. In each case the number of clicks, meaning jumps to the ground state, is close to the predicted value: 9 vs. 10 on the left and 73 vs. 70 on the right. 

{\it Energy uncertainty:}
The average $\langle E\rangle$ and the variance $\langle \delta E^2\rangle=\langle E^2\rangle-\langle E\rangle^2$ of the initial state $|\psi(0)\rangle$ are
\begin{eqnarray}\label{ave_var_1}
&&\langle E\rangle=\frac{\hbar\omega_{\rm Q}}{2}[1-2|\langle g|\psi(0)\rangle|^2]\nonumber\\&&\langle \delta E^2\rangle=(\hbar\omega_{\rm Q})^2|\langle g|\psi(0)\rangle|^2[1-|\langle g|\psi(0)\rangle|^2],
\end{eqnarray}
assuming eigen-energies $E_e=+\hbar\omega_{\rm Q}/2$ and $E_g=-\hbar\omega_{\rm Q}/2$.
We now compare expressions of Eq.~\eqref{ave_var_1} with the measurement outcome. We prepare the system $N$ times to the state of Eq.~\eqref{wave}, and measure the guardian photon each time. Assigning $N_g$ to be the number of observed click-down's and $N_e$ of click-up's, we have the expectation values for large $N$ as follows 
\begin{eqnarray}\label{aveE_1}
&&\langle E\rangle=\frac{N_g}{N}E_g+\frac{N_e}{N}E_e\nonumber\\&&\langle \delta E^2\rangle=\frac{N_g}{N}E_g^2+\frac{N_e}{N}E_e^2-(\frac{N_g}{N}E_g+\frac{N_e}{N}E_e)^2.
\end{eqnarray} 
Based on the previous discussion $N_g/N=P_{\rm click-down}=|\langle g|\psi(0)\rangle|^2$ and $N_e/N=P_{\rm click-up}=|\langle e|\psi(0)\rangle|^2$. Inserting these results to Eq. \eqref{aveE_1}, we again obtain Eq. \eqref{ave_var_1}, but now the "quantum uncertainty" of the initial state is transformed into statistical variance in the measurement results. Thus the statistics for the guardian photons follow the standard expectation of measurement outcomes in quantum mechanics. 
\begin{figure}
	\includegraphics[width=\columnwidth]{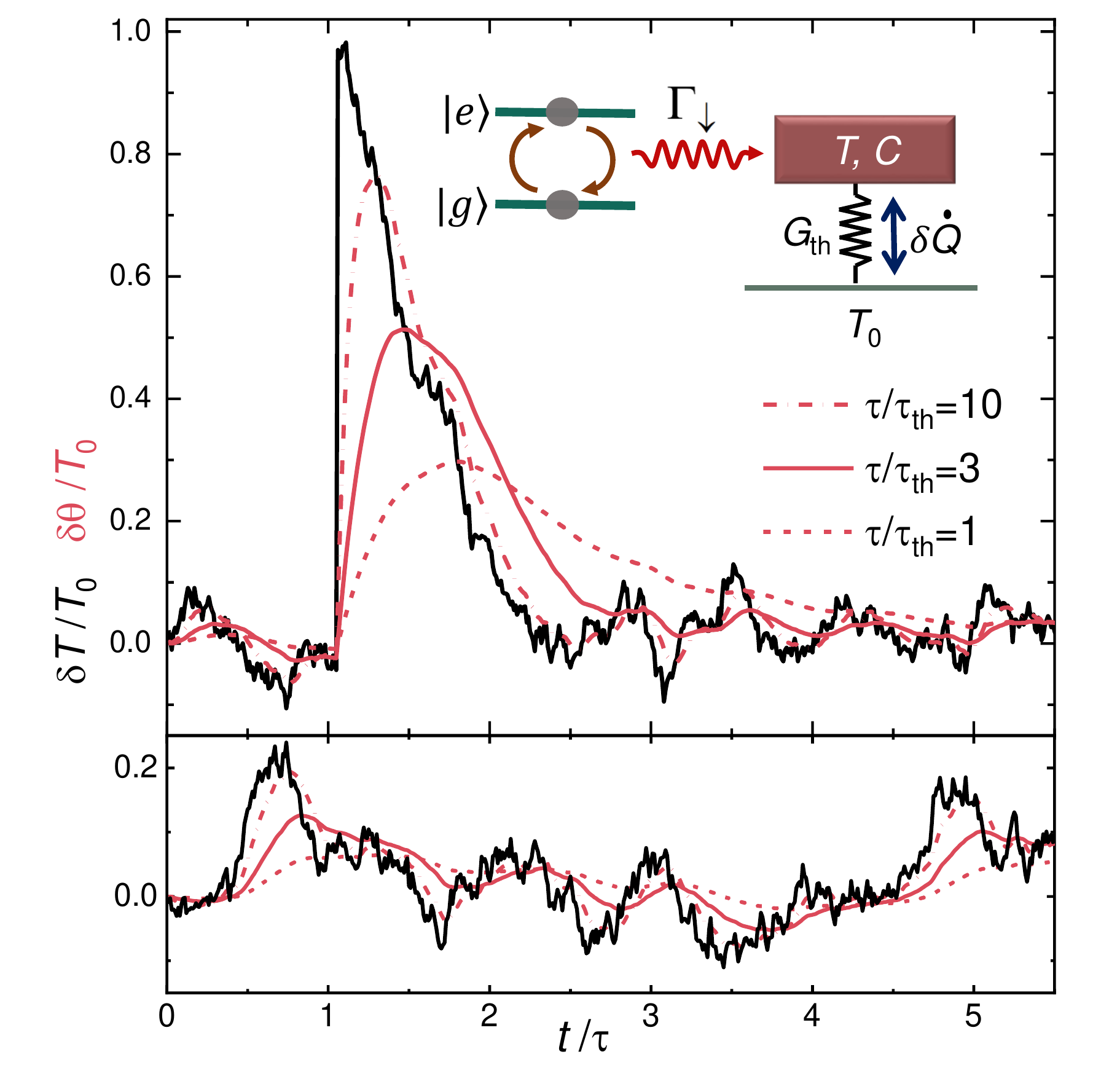}
	\caption{Expected response of a calorimeter. Time traces of absorber temperature based on qubit dynamics shown in Fig.~\ref{qubitbath} (a). In the simulations we use $\hbar\omega_{\rm Q}/k_{\rm B}T_0=100$ and $C/k_{\rm B}=100$. These parameters are for a copper absorber of $(0.1~{\rm \mu m})^3$ volume and $T_0=0.01$ K which are realistic based on recent experiments \cite{BKnoise,Anne}. In the top panel a jump occurs at $t\approx \tau$, clearly exceeding the noise level of equilibrium fluctuations (see text). In the lower panel no jump occurs in this time interval. The black solid lines show the exact absorber temperature for one realization of the experiment, and the red lines are outcomes of a noiseless measurement with three different response times $\tau_{\rm th}$ of the thermometer. Inset shows the model of the measurement set-up.}
	\label{Tt}
\end{figure}

{\it Temperature response of the absorber:} Based on our detecting scheme depicted in the inset of Fig.~\ref{Tt}, we can write the Langevin equation for the temperature of the absorber for small variations $\delta T=T-T_0$ as 
\begin{equation}\label{langevin}
C{\delta \dot T}(t)=-G_{\rm th}\delta T(t)+\delta \dot{Q}(t),
\end{equation}
where $C$ and $G_{\rm th}$ are the heat capacity of the absorber and the thermal conductance to the superbath, respectively, and $\delta \dot{Q}(t)$ is the instantaneous heat current on the absorber. The noise of the average heat current in this regime, $\delta\dot{Q}_{\rm ave}(t)=\frac{1}{\Delta t}\int_{t-\Delta t/2}^{t+\Delta t/2}dt' \delta \dot{Q}(t')$, over time interval $\Delta t$ is obtained as 
\begin{equation}\label{rms2}
\langle \delta\dot{Q}^2_{\rm ave}(t) \rangle=S_{\dot{Q}}(0)/\Delta t,
\end{equation}
where the low frequency spectral density is, according to the fluctuation dissipation theorem~\cite{FDT, fredrik} $S_{\dot{Q}}(0)=\int dt'\langle \delta\dot{Q}(t') \delta\dot{Q}(t'') \rangle = 2k_{\rm B}T_0^2G_{\rm th}$ in equilibrium. Introducing dimensionless time $u=t/\tau$ with $\tau=C/G_{\rm th}$ the thermal time constant, and discretizing it in steps $\Delta u = \Delta t/\tau$ leads to a coarse grained version of Eq. \eqref{langevin} as
\begin{equation}\label{dT}
\delta T(u+\Delta u)=(1- \Delta u) \delta T(u)+\sqrt{\frac{2k_{\rm B}T_0^2}{C}}\xi(u)\sqrt{\Delta u}.
\end{equation}
Here, we have normalized the noise as $\delta\dot{Q}_{\rm ave}(u)=\sqrt{\langle \delta\dot{Q}^2_{\rm ave} \rangle}\xi(u)$, where $\xi(u)$ has a Gaussian distribution with unit width $P(\xi)=\frac{1}{\sqrt{2\pi}}\exp (-\xi^2/2)$. The obtained results are not expected to depend explicitly on the value of the time step as long as $\Delta t \ll \tau$. 

Equation~\eqref{dT} forms the basis of Monte-Carlo simulations of temperature history of the absorber with $\xi(u)$ as the Gaussian distributed stochastic variable. On top of this evolution we add in Eq.~\eqref{langevin} the effect of stochastic energy absorption events $\delta \dot Q(t) =\pm \hbar\omega_{\rm Q} \delta(t-t_i)$ at the time $t_i$ of each quantum jump causing a sudden temperature change of the absorber with magnitude $\Delta T=\pm \hbar \omega_{\rm Q}/C$, where $+(-)$ refers to a qubit making a transition to $|g\rangle$ ($|e\rangle$). For low $T$ only the former transitions occur, as in Fig.~\ref{qubitbath}a. For numerical simulations (Fig.~\ref{Tt}), we assume a microwave photon with $\hbar\omega_{\rm Q}=k_{\rm B}\times 1$ K $=h\times 20$ GHz energy, a constant heat capacity $C/k_{\rm B}=100$ of the absorber, which is consistent with $C=\gamma \mathcal{V} T_0$, where $\gamma\sim 100$~${\rm Jm}^{-3}{\rm K}^{-2}$ for a typical metal, $\mathcal{V}=(0.1~\mu{\rm m})^3$, $T_0=10$ mK and $\Delta t=0.01\tau$. These numbers are based on assuming a superconducting qubit and a metallic resistor formed of the Fermi gas of about $10^8$ electrons with fast internal relaxation and weak coupling to the superbath via electron-phonon interaction, which are all experimentally feasible~\cite{ZBA,BKnoise,pothier}. We see in Fig.~\ref{Tt} that the signal-to-noise ratio for observing such a photon is about 10 under these conditions (top panel), which is consistent with our earlier estimates~\cite{BKnoise}. The time trace of the lower panel is a reference with no photon absorption. 

To model the actual temperature probe, we incorporate in the analysis its finite bandwidth. We do this by parametrizing it using a response time $\tau_{\rm th}$, such that the measured temperature $\theta(t)$ follows the actual temperature $T(t)$ calculated above via 
\begin{equation}\label{theta1}
\dot{\theta}(t)=-\tau_{\rm th}^{-1}(\theta(t)-\delta T(t)).
\end{equation}
Then with the time step $\Delta u$ we obtain
\begin{equation}\label{theta2}
\theta(u+\Delta u)=\theta(u)-\frac{\tau}{\tau_{\rm th}}(\theta(u)-\delta T(u))\Delta u.
\end{equation}
Naturally for $\tau/\tau_{\rm th}\gg 1$, $\theta(u)\simeq T(u)$, i.e, the thermometer follows the actual temperature, and for $\tau/\tau_{\rm th}\ll 1$, $\theta(u)={\rm const.}$, meaning that it does not respond to the changes of $T$. Figure~\ref{Tt} shows numerical results of $\theta(t)$ with a few values of $\tau/\tau_{\rm th}$. If one were to consider the noise of the thermometer itself, one could add a Langevin term to Eq. \eqref{theta1} with proper noise characteristics, but we feel including this would be beyond the scope of this paper in the absence of actual experimental data. Finally, we note that the results can be generalized to calorimetric fluorescence detection acting on an arbitrary quantum system. In particular the measurement of single emitted photons as described in the previous paragraph and in Fig.~\ref{Tt} stay unaltered. The calorimeter thus presents a continuously operating detector capable of registering the clicks due to single photon events, with the additional bonus of being able to measure (linearly) the energy of the quanta.  

In summary, we have presented a model for a calorimetric fluorescence measurement of an open quantum system. We use the stochastic quantum trajectory theory and verify that it is consistent with the common expectations of measurements in quantum mechanics. We demonstrate explicitly that quantum thermodynamic measurements of superconducting circuits are possible down to single quantum level with a realistic continuously operating wide-band detector at a sufficiently low temperature. 

We acknowledge Klaus M\o{}lmer for sharing his knowledge about the topic and for valuable discussions. We thank Yasuhiro Utsumi for useful comments on the manuscript. This work was funded through Academy of Finland grant 312057 and from the European Union's Horizon 2020 research and innovation programme under the European Research Council (ERC) programme and Marie Sklodowska-Curie actions (grant agreements 742559 and 766025).

\end{document}